# An XR rapid prototyping framework for interoperability across the reality spectrum

Efstratios Geronikolakis[1,2], George Papagiannakis[1,2]

[1] Foundation for Research and Technology Hellas, 100 N. Plastira Street, 70013 Heraklion, Greece

[2] Computer Science Department, University of Crete, Voutes Campus, 70013 Heraklion, Greece

**Abstract** Applications of the Extended Reality (XR) spectrum, a superset of Mixed, Augmented and Virtual Reality, are gaining prominence and can be employed in a variety of areas, such as virtual museums. Examples can be found in the areas of education, cultural heritage, health/treatment, entertainment, marketing, and more. The majority of computer graphics applications nowadays are used to operate only in one of the above realities. The lack of applications across the XR spectrum is a real shortcoming. There are many advantages resulting from this problem's solution. Firstly, releasing an application across the XR spectrum could contribute in discovering its most suitable reality. Moreover, an application could be more immersive within a particular reality, depending on its context. Furthermore, its availability increases to a broader range of users. For instance, if an application is released both in Virtual and Augmented Reality, it is accessible to users that may lack the possession of a VR headset, but not of a mobile AR device. The question that arises at this point, would be "Is it possible for a full s/w application stack to be converted across XR without sacrificing UI/UX in a semi-automatic way?". It may be quite difficult, depending on the architecture and application implementation. Most companies nowadays support only one reality, due to their lack of UI/UX software architecture or resources to support the complete XR spectrum. In this work, we present an "automatic reality transition" in the context of virtual museum applications. We propose a development framework, which will automatically allow this XR transition. This framework transforms any XR project into different realities such as Augmented or Virtual. It also reduces the development time while increasing the XR availability of 3D applications, encouraging developers to release applications across the XR spectrum.

**Keywords: Virtual Reality, Augmented Reality, Mixed Reality, Extended Reality, Cultural Heritage, Virtual Museums**

# 1 Introduction

Throughout the years, computer graphics applications are becoming more and more prevalent each day. People use them for a variety of purposes, for instance, entertainment [23], education [31], training [19], even research [19].

A crucial term mainly mentioned in this work is XR. Extended Reality (XR) is a fusion of all the realities – including Augmented Reality (AR), Virtual Reality (VR), and Mixed Reality (MR) – which consists of technology-mediated experiences enabled via a wide spectrum of hardware and software, including sensory interfaces, applications, and infrastructures. XR is often referred to as immersive video content, enhanced media experiences, as well as interactive and multi-dimensional human experiences. [30]

The number of 3D applications that exist across the XR spectrum is considerably limited. Discovering a Virtual Reality application for instance, that exists in Augmented Reality as well, is a rare situation. It is comprehensible, considering that the two previously mentioned realities greatly differ in terms of user experience (interaction). As a result, the amount of effort needed for transitioning such an application from Virtual Reality to Augmented Reality or vice-versa (across the XR spectrum) would be approximately identical to that of developing it from scratch. This fact is the primary reason why developers do not consider releasing an application across the XR spectrum. It is quite unfortunate since there is always a chance for it to be more appealing in another reality than the one it was initially designed for.

Depending on the application content and its general purpose, it might be more suitable for Virtual Reality than Augmented Reality, for instance. Such an example would be surgical training applications [20]. In such cases, we need the users to fully immerse themselves and try to remember the steps and movements (with the controllers) they executed inside the virtual world. That would not be possible for a mobile device with Augmented Reality.

The previously mentioned problems would be addressed if the transitioning procedure across the XR spectrum was quick and could be executed effortlessly. Developers would quickly port their application through the XR spectrum and discover the most appropriate case for each application. This problem can be addressed by developing a framework, able to be used through the game engine, which would automatically transit the application project across the XR spectrum (generating a ready-to-operate project), while respecting the developers' choices.

This work aims to relieve the developers (and especially new ones) from a burden they regularly face: the transition of a computer graphics application across XR (from VR to AR, for instance) and more specifically, virtual museum ones. This transition usually requires many actions from the developers' side. Of course, tutorials that exist online only explain how to start an application in the chosen reality from scratch, as shown in [2]. Similar are the cases for the other platforms/realities as well. Our tool aims to relieve developers from this confusion and offer them a swift solution to their problem, which will be an operating project for the reality of their choice, for their application.

Another intention of this work is to reveal the way that a cultural heritage application can transcend across XR, (meaning adapting the needs of the application to the new reality/device requirements) by using our framework. It solves the problem above and is a part of the results of this work. An example of such application XR transition is described in [7], where an application transits from Virtual Reality to Augmented Reality manually. Our framework automates an important part of this process.

We split this work into five sections. Broadly, this dissertation's main roadmap is the introduction, previous work, the definition of the problem, and its solution alongside its evaluation.

In the second section, we mention some other inspiring cultural heritage applications and some authoring tools that contribute to the development of such applications. Following that, we discuss some applications that exist across XR, which is what we designed our tool to do automatically. Then, we cite other XR applications, for which their primary purpose was to educate the users (serious games).

In the third section, we introduce the essential elements of each of the platforms and devices used for this work. Afterwards, we declare the definition of the principal problem of this work. Then, we present our framework, the "XR transition manager", which performs a fast and easy transition across XR for an application. In the fourth section, we present our framework's evaluation scheme, along with the results, and we draw some conclusions. Lastly, in the fifth section, we discuss the future work we could do, in order to improve our tool further.

## 2 Previous Work

There are several types of work that we will present in this section. Most of them concern the preservation of Cultural Heritage because our framework is targeted towards the transition of virtual museum applications across XR. We will present some exciting state of the art virtual museum applications. As our framework will also involve the ability of transition across XR, we will examine some existing authoring tools and applications that work in various realities.

### 2.1 Virtual museums

Nowadays, virtual museums have become more and more prevalent. The idea of designing a virtual version of a real museum is a fascinating one, as it is a way to contribute to cultural heritage preservation. That is because, by generating a virtual museum application and distributing it to people, they can gain access to the virtual version of the particular museum from anywhere in the world. It is a very innovative idea since every person that maintains a mobile device will virtually visit the museum of their desire, without having to travel directly to it, making their life more convenient. A rather extensive study [18] presents different Mixed Reality technologies from various researchers regarding the area of virtual museums.

### 2.1.1 Holographic Virtual Museums

Since the advancement of holographic technology, AR headsets are evolving, including interactive features like gesture and voice recognition and improvements on resolution and FOV. Besides, untethered AR headsets paved the way for mobile experiences without external processing power from a PC. Such embedded systems facilitate excellent tools to represent virtual museums [14] due to their lack of cables and enhanced interactive capabilities. Virtual Museums are institutional centers in society's service, open to the public for acquiring and exhibiting the tangible and intangible heritage of humanity for education, study, and enjoyment. Also, True Augmented Reality has recently been defined as a modification of the user's perception of their surroundings that the user cannot detect [24] due to their realism. Virtual characters and objects should blend with their surroundings, attaining the "suspension of disbelief".

Many approaches on holographic cultural heritage applications emerged in modern years, each one concentrating on a different aspect of representing the holographic exhibits within the real environment. A published survey [12] investigated the impact of Virtual and Augmented Reality on museums' overall visitor experience, highlighting the social presence of AR environments. [17] presented a correlation of the latest methods for the rapid reconstruction of real humans using as input RGB and RGB-D images. They also propose a complete pipeline to compose highly realistic reconstructions of virtual characters and digital asserts suitable for VR and AR applications. Storytelling, Presence, and Gamification are three critical fields that should be considered when creating an XR application for cultural heritage. [18] presented a comparison of existing MR methods for virtual museums and pointed out the importance of these three fields for applications that contribute to cultural heritage preservation [11]. Furthermore, in [10], fundamental elements for MR applications alongside examples are presented.

Another recent example [31] introduced two Mixed Reality Serious Games in VR and AR, comparing the two technologies over their capabilities and design principles. Both applications showcased Knossos's ancient palace in Minoan Crete, Greece, through interactive mini-games and a virtual/holographic tour of the archaeological site using Meta AR glasses. [1] successfully published an AR application for visualizing restored ancient artefacts based on an algorithm

## 2.2 Authoring tools for content creation in MR

Content is an essential part of a 3D application (if not the most crucial one). Extensive care must be taken during content development, as it constitutes the application and user experience base. Since creating content is a time-consuming and challenging procedure, different authoring tools for content creation have come to the surface to ease developers' lives. In this section, we present some of the prevailing works in this field.

### 2.2.1 Platforms for Gamified Content Creation

Authoring tools and additional content creation platforms emerged in modern years to fulfill the demand for interactive MR applications. RadEd [21] highlights a new web-based teaching framework with an integrated smart editor to create case-based exercises for image interaction, such as taking measurements, attaching labels, and selecting specific parts of the image. It facilitates a framework as an additional tool in complex training courses like radiology. BricklAyeR [26] is a collaborative platform designed for users with limited programming skills that allows the creation of Intelligent Environments within a building-block interface. In [3], the authors present tools for interactive virtual human population in the concepts of cultural heritage. These tools utilize open source generic frameworks for generating interactive 3D virtual environments and content. ExProtoVAR [22] is a lightweight tool to produce interactive virtual prototypes of AR applications designed for non-programmers lacking AR interfaces experience. ARTIST [13] is a platform, which provides methods and tools for real-time interaction between human and non-human characters to generate reusable, low cost, and optimized MR experiences. It aims to develop a code-free system to deploy and implement MR content while using data from heterogeneous resources semantically. The aforementioned solutions provide developing environments to generate MR experiences. However, they lack advanced authoring tools and educational curriculum to support advanced educational - training scenarios. Lastly, in [33], the authors propose a gamified way of content creation for a training application, through a user interface by connecting blocks of events or setting up the desired events through Virtual Reality.

### 2.2.2 Unity MARS

Unity MARS [28] is a novel AR authoring tool, produced by Unity3D in 2020. It is unique since it offers the ability to develop and test AR applications from the Unity MARS environment without building the application each time. With the use of proxies, which are 3D objects imitating real-world objects, and "fuzzy authoring" developers can define the minimum and maximum measurements for them rather than code precise values. Moreover, with a relatively simple drag-and-drop feature, developers can place their 3D models in the scene, and Unity MARS produces all the appropriate proxies and conditions for them. It also supports different kinds of real-world data, such as images and surfaces. In the near future, it will also support body tracking. It is a fascinating and time-saving authoring tool for AR that aims to speed up AR development.

## 2.3 Applications existing across XR

Following a rather extensive search for previous works on applications existing across XR, we did not find any scientific results. It is acceptable to the perspective that it is rather challenging for an application to exist across XR. From our experience, we believe that one of the reasons for this absence of examples could be the porting complexity. Of course, it is a procedure, which is not unachievable. It is doable, but since it is a time-consuming procedure [9], many people refrain from commencing it. They prefer generating new content for a particular platform and push on. They do not ponder porting the same applications to other platforms, since they may consider this "recycling" of the same application. All the previously mentioned thoughts are based on our opinion, of course.

Another reason could be that apart from the challenge of the porting procedure, each platform has its specifications and requirements. As a result, an application in Virtual Reality that uses controllers would need total rework to operate in a mobile device and Augmented Reality. Mobile devices do not encourage the use of controllers. Their primary input device consists of a touchscreen. This fact completely changes the whole user experience, and it is reasonable to refrain the developers from attempting the port.

Multiple applications have certain principles and require a specific structure in order to offer their full experience to users. Suppose one of these principles is absent (in our example, the fact that users should be able to move their hands around freely and see them in the application). In that case, the immersion and, as a result, the user experience drops significantly.

We managed to spot one example, nevertheless. That would be a Virtual Reality application, which exists in Mixed Reality as well (HoloLens). Within this application, users perform a surgical operation, named Total Knee Arthroplasty. More specifically, users follow a sequence of actions, visualized by holograms, to complete the operation. We found out that an approach to transfer this application to HoloLens exists. However, it remained incomplete (it does not contain the whole operation), probably because of the reasons we explained.

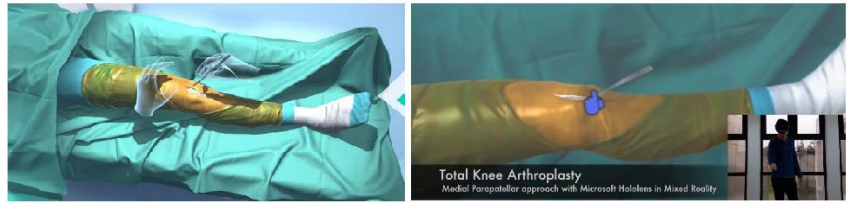

**Figure 1.** Total Knee Arthroplasty in Virtual Reality (left) and Mixed Reality (right).

### 2.4 Mixed reality serious games for smart education

Considering we consumed a large amount of time to produce our virtual museum application and worked to make it available for multiple platforms, we scrutinized for work regarding Mixed Reality serious games. We performed this research to determine the gamification methods that were used by researchers in their respective works, to collect ideas for our virtual museum.

In [32], the authors designed an application related to Knossos archaeological site. This application intends to educate users concerning the history of Knossos through a series of mini-games. They created several versions of this application across XR. They perceived that the same application could not be applied across XR because each one had its boundaries and characteristics. In [15], they clearly define the meaning and significance of serious games and gamification while presenting different technologies for immersive heritage applications. Another work, bearing valuable information for constructing Mixed Reality applications for cultural heritage, is specified in [5]. Similarly, in [16], the authors present an inspiring, serious game application for intangible cultural heritage and, more explicitly, dancing. Users can learn many traditional dances by following a sequence of movements, while the software provides them with feedback regarding how well they performed these movements.

### 3 Our Methodology – Contribution

Each XR application has different essential elements regarding the camera and user interaction functionality. These elements vary depending on the platform and hardware that the application operates. The platforms/operating systems that most 3D applications use (in Unity3D [27]), and we studied in this work are four. The first one is "PC, Mac & Linux Standalone", for applications that operate on a desktop computer. The second one is the "Android" platform for applications that operate on a mobile device or headset that supports android. Next is "iOS" for mobile devices that run Apple's iOS and, lastly, "Universal Windows Platform" for mobile devices utilizing Windows. The hardware on which most 3D applications operate is either VR Head Mounted Displays (HMDs) or mobile devices (smartphones, tablets, and more).

### 3.1 Virtual Reality Applications

Virtual Reality technology generates an artificial environment that immerses the users, causing them to believe that they are a part of it, doing there and being there. Although this technology has emerged long ago, it is in the latest years that it made its first steps in the market and became well-known. For the users to enter the virtual world that a VR application provides, they have to wear a VR head-mounted display (HMD), which precludes them from having access to the real world, as long as they wear it. As a result, users experience full immersion. This technology is used widely for many purposes, from entertainment [23] to training [19] and research [19].

### 3.2 Augmented Reality Applications

Augmented Reality technology merges the real with the virtual world. Applications made utilizing this technology usually operate on mobile devices containing at least one camera component. A camera is profoundly needed because this is the users' "window" to the virtual world, where the virtual 3D objects will reside. Augmented Reality offers partial immersion since users still have access to the real world. Both iOS and Android mobile devices are eligible for operating Augmented Reality applications. Each platform has produced its version of Augmented Reality Software Development Kit (SDK), namely ARKit [4] and ARCore [8].

### 3.3 Holographic Augmented Reality

Holographic Augmented Reality, as the name implies, is very close to straightforward Augmented Reality. It differs in the way that all virtual objects are rendered as holograms. With Holographic Augmented Reality's aid, users can observe these holograms through a special HMD that highly increases realism. As a result, not only do they experience the presence of virtual objects being in their room, but they can also interact with them using hand gestures. Thus, Holographic Augmented Reality is bound to partial immersion, since users have access to the real world, but in this case, realism is even higher compared to straightforward Augmented Reality.

### 3.4 All-in-one Unity XR transition manager

Transitioning across XR, or even setting up a new project for the first time may be time-consuming for developers [9]. For instance, assume that a developer desires to start producing an Augmented Reality application for Android mobile devices. After starting Unity, they will have to ponder some crucial elements:

1. Is there a specific Software Development Kit (SDK) for this project that could be useful?
2. Where can this SDK be downloaded?
3. Is it open-source/free?
4. Which version is the most appropriate? (Mostly, the latest versions are the best choice, but there are also some other situations like testing or experimenting, where an older SDK version is required.)

These are the main questions that cross the developer's mind. When the developer finishes thinking about the above, what they need to do is probably open a web page and begin searching one-by-one the above questions to attain the required answers. Sometimes this could be a relatively fast procedure (especially if the developer has much experience). However, in the case of newcomers de-velopers, this could be quite a time-consuming procedure [9]. It is where we can prove our work to be beneficial both to experienced and inexperienced developers. We describe the first part of it below.

## 3.5 XR transition download manager

Another question that arises is, "What if the SDK searching and downloading procedure took place from inside Unity?". It is why we created an SDK download manager, which is accessible through the game engine.

### 3.5.1 The basic architecture of XR Transition manager

The "XR Transition manager" has a simple architecture. Its primary function is to connect to the main hosting website of the desired SDK. If the manager detects the requested version, it downloads the selected SDK through that connection and then follows the XR transition procedure. The manager handles the camera component and sets it up correctly. Developers need to specify the game-object of the camera, in any case. Overall, the manager connects with the following four components:

1. Websites hosting the SDKs.
2. Downloaded SDKs.
3. Unity scene.
4. Main camera component of the scene.

We can view the central architecture of the manager in Figure 2 below.

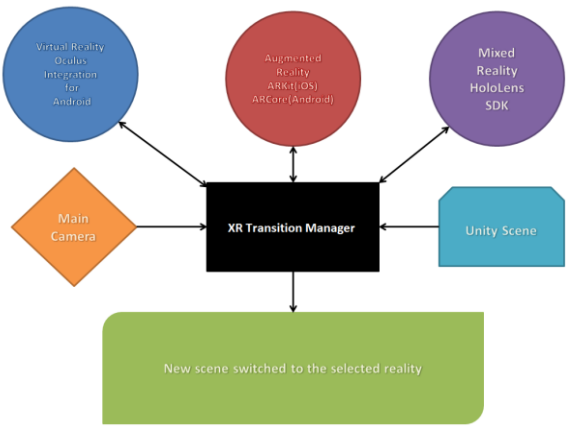

**Figure 2.** The basic architecture of the XR Transition manager.

### 3.5.2 The main code structure and logic of XR Transition Manager

"XR Transition Manager" consists of a combination of editor scripts. We designed them using the C# programming language. We divided them into three groups, namely the "Menu Script", "SDK Download & Setup Scripts" and "XR Transition Scripts". We present these groups in detail below.

#### 3.5.2.1 Menu Script

This collection consists of one script only, namely RealitiesMenu. This script is the backbone of the whole framework. It is responsible for calling and invoking the function that developers select when they click an option from the respective Unity3D menu. For instance, if developers navigate to Realities menu and click the "ARCore" option under Switch Reality/Augmented Reality, the base script is responsible for calling the function, which contains all the necessary steps to perform an XR transition.

#### 3.5.2.2 SDK Download & Setup Scripts

In this collection, the SDK downloading and installing scripts exist. There are four scripts in this category, one for each of the currently supported SDKs. There is one for ARKit (DownloadARKitWindow), one for ARCore (DownloadARCoreWindow), one for Mixed Reality (DownloadMixedRealityWindow), and another one for Oculus Integration (DownloadOculusWindow). All these scripts have the word "Window" appended to their name, considering the very first thing they do is to enable a window, in which developers will be able to select their desired version to install. In addition, they can check whether they would like the downloaded SDK to be installed immediately after downloading finishes from this window.

These scripts accommodate an OnGUI function, which contains everything drawn to the window from which developers select the SDK version that interests them. As its name informs us, it contains everything that exists On the Graphical User Interface.

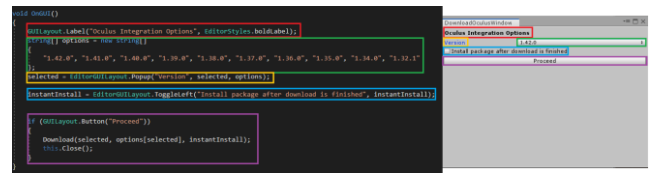

**Figure 3.** The contents of the OnGUI function with their respective matches to the window that the developer sees. (Each color on the left matches with its respective on the right.)

The OnGUI function follows the Download function, where the main downloading procedure takes place. Within it, there are some hardcoded links, each representing a version of the preferred SDK. Depending on the developers' SDK version decision, the script selects the appropriate link. The downloading procedure begins once the developers' internet connection is verified; otherwise, the script prints a representative error message. During downloading, we calculate and display the progress using a progress bar window. We also give developers an option to cancel the download procedure by clicking the "Cancel" button located at the window's underside.

When downloading terminates, we save the downloaded data inside the project assets folder. Then, the "Download"

function proceeds by checking whether developers chose to install the SDK instantly or not. If they chose to do so, it imports the package, using the default package importing method of Unity3D. However, the same does not hold for ARKit, where the SDK comes in compressed zip files rather than unitypackage files. For ARKit, we take extra steps to decompress the file and install it, since Unity3D does not offer a built-in system to decompress zip files automatically. We decided to use a third-party decompression tool, termed UniZip [29], which handles the specific procedure. Once the installation step finishes, the definitions section takes place.

We take particular precautions for scripts containing code and expressions targeting specific SDKs because they will not compile unless these SDKs exist in the project. For instance, scripts that belong to the "XR Transition Scripts" collection include specific commands and utilize prefabs existing in the SDKs they represent. If these SDKs are not present during the compilation phase, it will result in compilation errors, prohibiting the whole project from operating. Thus, we created some specific definitions (one for each SDK) for the framework to identify the currently installed SDK each time. Of course, this applies only to SDKs that users installed through our framework. Using these definitions, the code that will be compiled each time will be the one that represents the currently installed SDK, thus solving the compilation problem for SDK specific code.

```
RemoveDefineIfNecessary("ARCORE_SDK", BuildTargetGroup.Standalone);
RemoveDefineIfNecessary("ARCORE_SDK", BuildTargetGroup.iOS);
RemoveDefineIfNecessary("OCULUS_SDK", BuildTargetGroup.Standalone);
RemoveDefineIfNecessary("OCULUS_SDK", BuildTargetGroup.Android);

AddDefineIfNecessary("ARKIT_SDK", BuildTargetGroup.Standalone);
AddDefineIfNecessary("ARKIT_SDK", BuildTargetGroup.iOS);
```

**Figure 4.** An example of managing definitions for the case of ARKit.

#### 3.5.2.3 XR Transition Scripts

"XR Transition scripts" is the final collection of the "XR Transition manager". In this collection, we added four scripts, each of them being responsible for transitioning to one of the four supported platforms. Once again, the OnGUI function exists in each, setting up the window, where the developers select the main camera. Once developers press the proceed button, the script applies all the required options in the player settings menu. Then it performs a platform switch to the desired platform, constructs, and instantiates all necessary prefabs to generate a scene, which will operate successfully for the desired platform. Of course, we take some extra precautions to ensure that the manager will continue operating even if something does not go as intended. For instance, it does not take for granted that the correct SDK is installed even in this case. On the contrary, it tries to make sure that everything regarding the SDK is installed and exist. Otherwise, it interrupts the procedure by notifying the developers through an error message.

### 3.5.3 Downloading and installing SDKs through the XR Transition Manager

Our manager consists of editor scripts, written in C# that form a new menu, in the default Unity menu bar named "Realities". A dropdown appears by pressing this button (Fig. 5), and one of the options contained inside is "Software Development Kits". Another menu appears on the right by clicking this option, containing all available free SDKs depending on the developers' desired platform. We currently support ARCore for Android mobile devices, ARKit for iOS devices, Oculus integration for Android VR (Oculus Go/Quest), or Windows PC (Oculus Rift). Finally, Mixed Reality toolkit for Windows Mixed Reality headsets as well as Microsoft HoloLens.

By clicking the required SDK, a window appears, granting developers a choice between different versions of this SDK (Fig. 6). Developers are offered to pick from a dropdown menu the preferred version. Also, they are prompted to click on a checkbox, in case they require the SDK to be imported/installed right after it finishes downloading. When they set all the preferences, they must click the "Proceed" button to continue.

An internet connection is vital in order to download the selected SDK. The download speed depends on the developers' internet line capabilities. Though they will experience the same speed, they would if they downloaded it from the original site.

### 3.6 Importance of the XR Transition SDKs download manager

As we mentioned previously, the download manager could be handy for developers while starting a new project. It speeds up the procedure and makes it exceptionally effortless. Also, it increases productivity, since it reduces the context switch between the current workspace and browsing [25] that developers may result to in case they do not remember all the necessary settings for the platform of their choice. Without our manager, developers, apart from searching and downloading the desired SDK on their own, would also have to work further to import it (unzipping the SDK and placing it in the right place). Our download manager handles this likewise. Additionally, it can be useful when developers would like to transit their application across XR. It is able to download the new SDK and get the developers up and running to develop their application in no time.

### 3.7 XR Transition Feature

Our "XR Transition manager" includes another fundamental and critical functionality. It downloads whichever supported AR/VR/MR SDK needed, which saves a substantial amount of developers' time. However, even though an SDK is downloaded and installed, further work is required from the developers' side for it to work correctly for the final application to be built and operate flawlessly. For instance, to build one that uses Google's ARCore, the following options must be set:

- The target platform must be Android.
- The default Graphics API must be either OPENGLES2 or OPENGLES3.
- The packages "AR Foundation", "Multiplayer HLAPI" and "XR Legacy Input Helper" must be installed in the current project.
- The minimum Android version must be 7.0 (API Level 24).
- The Unity setting "ARCore supported" must be enabled.
- Search in the SDK folders to find the specific camera default prefab and place it in the scene.

Developers must memorize all the above to set the project configuration correctly and build a sample application. It can be challenging since all these steps contain specific details (such as package names and settings) that are hard to retain. Especially for new developers, this can be rather frustrating. Usually, developers open a web browser and begin searching to find the previously mentioned settings. If they are fortunate enough, they find them in a few seconds, but there is a possibility that the correct website will not appear to them immediately. It is why our manager includes an "XR Transition" feature.

Once developers have downloaded and installed an SDK through our manager, they can decide to perform XR transition through Realities menu. Specifically, by pressing the Realities button and then the Switch Reality one, they can select the reality of their desire (Fig. 9). When they define it, another window appears, prompting them to define their main camera game-object present in their scene (Fig. 10). In a 3D application, a camera object is mandatory for the player/user to view the scene and navigate. Thus, developers have to locate their main camera object in the Unity scene and place it in the appropriate box located in the window mentioned before. Once they do so, they have to click the "Proceed" button, and the SDK manager takes care of everything else. Specifically, it switches to the appropriate target platform, installs the required packages, sets up the project correctly, and instantiates the appropriate camera prefab in the scene, in the same position and rotation as the previous camera object. The latter is deactivated since it is "replaced" by the new one, but it remains in the scene, in case developers would like to do something else with it, or save it if they would like to return to the previous reality. This procedure exists in our SDK manager for switching to Augmented Reality (Android and iOS), Virtual Reality (Mobile VR – Oculus Go/Quest), and Mixed Reality (Microsoft HoloLens).

### 3.8 Using the XR Transition Manager

"XR Transition Manager" is a framework for Unity3D that aims to speed up the developing process of 3D applications. In section 3.11, we described our framework's primary technology. In the next sections, we will present some basic examples regarding its usage.

### 3.9 Downloading and installing an SDK

Once developers import all the required manager scripts correctly in their project, a menu "Realities" should appear in the menu bar of Unity3D. This menu contains four options. The first one is "Software Development Kits" and provides another four should the developers hover the mouse over it. These options are "Download ARCore", "Download ARKit", "Download Oculus Integration" and "Download Mixed Reality Toolkit".

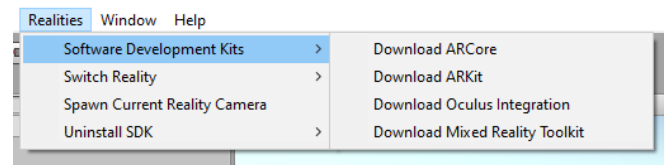

**Figure 5.** The supported SDKs, as they are presented in the "XR Transition Manager" menu.

Currently, our manager supports these four SDKs. By selecting one of these, developers can download the corresponding SDK. In this section, we will highlight the case of ARCore. However, the same procedure applies to the rest of the SDKs. When we click the "Download ARCore" option, the following window appears.

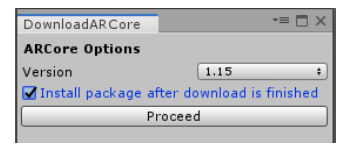

**Figure 6.** Choosing the version of ARCore and its installation before downloading it.

This window provides developers with the following options:

- **Version:** Developers have the option to select the SDK version of their choice. The versions are currently hardcoded in the manager. Therefore, to add newer versions, a new version of our manager has to be released as well each time.

- **Install package:** If developers check this button, the SDK is imported and installed automatically after download finishes. Otherwise, either a .unitypackage or a .zip file will appear in the project path for the developer to install/import it manually.

When developers set the version and the package install option, the procedure continues by clicking the "Proceed" button.

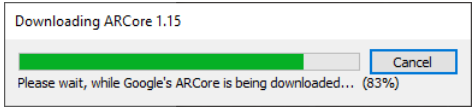

**Figure 7.** The progress of SDK downloading.

The window depicted above indicates the progress of the download procedure. Developers have also the option to stop it at any time by clicking the cancel button. In case of success, the following window appears, indicating that the download procedure was successful.

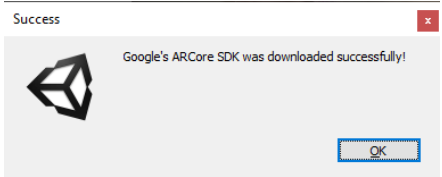

**Figure 8.** Success message for the successful download of SDK.

After developers click the OK button, depending on their initial choice, the SDK will be imported and installed, or they will receive a .unitypackage file containing the SDK, to install it by themselves.

### 3.10 Switching reality

The second option that "Realities" menu provides is transitioning the project across XR depending on the developers' choice. Again, in this example, we will focus on ARCore, but the same procedure applies across XR. As visible from the image below, there are three options under the "Switch Reality" tab.

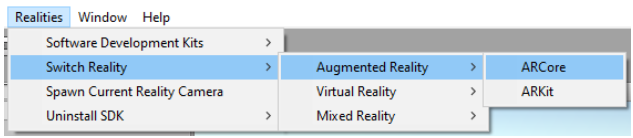

**Figure 9.** The supported realities of switching to, depending on the selected SDK/platform.

- Augmented Reality: This reality is mostly for mobile devices equipped with a camera. It is because, as mentioned previously, Augmented Reality offers partial immersion. For Augmented Reality, the two sub-options that are currently supported by our manager are ARCore and ARKit.

- Virtual Reality: This reality is for an application operating on a Head-Mounted-Display (HMD). Currently, our manager supports only Oculus Integration, rendering the supported devices to be mobile VR and Oculus Go/Quest.

- Mixed Reality: This reality is for VR/AR applications operating on Microsoft devices. Currently, we tested this option on Microsoft HoloLens (Holographic Augmented Reality).

Developers should consider switching to a scene supporting the SDK that they have downloaded. If they have not downloaded the correct SDK, an error message will notify them about it, and the manager will prompt them to download it.

When developers select the reality, they would like their project to switch into, the following window appears.

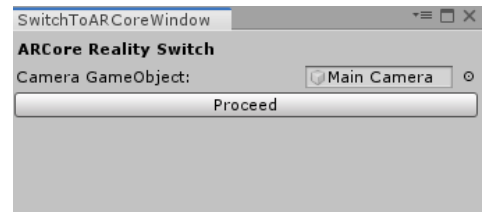

**Figure 10.** Settings window of switching to Augmented Reality for ARCore/Android.

The next step is to specify the main camera of their application, which serves as the users' eyes. Once it is set and developers click the "Proceed" button, the procedure of XR transitioning begins. The time developers have to wait afterward depends on the size of their project. This action performs a platform switch, sets all the required settings for the current SDK, and adds the main prefabs/objects in the scene required by the SDK for a successful and working build. Of course, after the operation finishes, developers can add/delete SDK related prefabs according to their desire. The XR transitioning procedure performs all the necessary options in a few seconds; otherwise, the developers would have to do it manually.

### 3.11 Uninstalling an SDK

The uninstalling feature could not have been missing from our SDK manager. There is a separate option in the "Realities" menu regarding uninstalling an already downloaded SDK. The image below depicts it.

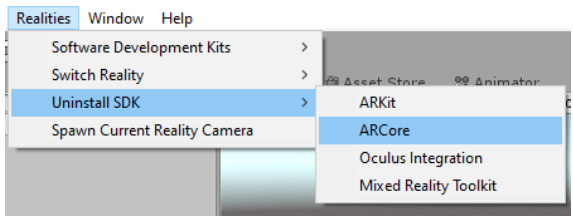

**Figure 11.** The SDK uninstall feature.

Someone could submit the following question. "Why should a developer do that through the 'Realities' menu and not by just selecting the SDK folder and manually deleting it?". That would be the same as deleting an unwanted program's folder in our computers. The program would not be there, but a lot of information and data would remain internally. The same holds for our SDK manager. Developers may delete the SDK folder separately, but all the options and settings our manager performed during this SDK installation would remain. These additional settings reset by our manager are the main reason for developers to have a "healthy" and "ready to operate" project after the SDK removal.

### 3.12 Automatic reality working camera

Sometimes, developers would like to place a working camera on their current project and reality, without needing to undergo an XR transitioning procedure (testing purposes, for instance). Our manager handles this case since it offers an option to spawn an XR specific camera on demand. We call this option "Spawn Current Reality Camera". When selected, our manager automatically detects the platform and installed SDK developers are working on and adds the corresponding test-ready camera to the scene.

### 3.13 Development of XR Transition Manager

The development procedure of the "XR Transition Manager" was relatively smooth, with a few challenges that took place along the way.

#### 3.13.1 The SDK handling of XR Transition Manager

The "XR Transition Manager's" central idea is to host several valid download links to each version of various SDKs and provide functionalities of installing and setting them up. It aims to ease the developers to quickly develop and build their applications without any SDK related errors. For each SDK, we include links to its stable versions in the code used for downloading it, in case the developers select the specific version. We use the UnityWebRequest API to validate the connection and proceed to the actual download. When it finishes, the downloaded file is recorded temporarily in the assets folder to be imported and installed to the project. We import the file using the ImportPackage utility of Unity. Unity offers the ability to make a prompt appear to developers when importing a package, enumerating the package contents to specify which components to install. We decided to disable this prompt since we install only the SDK's mandatory components (thus, everything is needed) to save time during installation and importing. After these procedures end, we delete the temporary downloaded file, leaving only the installed components.

#### 3.13.2 Challenges Faced During Development

The main challenges that we faced during the development of the "XR Transition Manager" were two:

Firstly, Apple's ARKit SDK is distributed in .zip form, instead of .unitypackage. However, the ImportPackage utility of Unity would not be successful in installing such files. To overcome this issue, we had to install Apple's ARKit without using the utilities of Unity. We investigated different approaches by writing our C# functions but to no avail. Eventually, we found a third-party Unity plugin called UnityZip; we imported it to the project and used this to decompress .zip files.

Secondly, another challenge that we faced was downloading the SDK packages. In the beginning, we considered downloading all the versions of each SDK, save them locally (as part of our manager), and install them, reducing the download time to zero. In the end, this approach did not produce any favorable results, as some SDKs were quite large (the size of Oculus SDK consists of some hundreds of MBs and has many versions), thus forming a space bottleneck. Moreover, this would not have been practical, as it would not offer portability. So, we tried experimenting with different ways to download these SDKs through C# code. We then discovered that using UnityWebRequest was our best bet since it was easy to utilize and practical. However, when we used it, we could not find the downloaded file. It was another challenge for us since we believed that this utility saved the file after downloading it. After some experimenting and searching, we found out that the file needed an extra step to save it successfully. We simply had to check the download operation and find out when it finishes. Then we had to retrieve the downloaded data from the web request structure and save it to our desired path using the File utility.

## 4 Results and Conclusions

In this section, we will summarize our work. It was a fascinating topic, which we enjoyed working on. Since there is no actual limit to what can be done or added to this work to make it even more helpful and meaningful to the developers, we will discuss some potential ideas to be added in the future.

### 4.1 Summary

In this work, we initially presented a framework for manually transitioning a cultural heritage application across XR (from VR to holographic AR). We presented how we trans-ferred a Virtual Reality application to holographic Augmented Reality and the challenges we faced during this platform switch. Afterward, we introduced the main problem with application transitioning across XR in a 3D application and discussed why this should occur automatically. We then described "XR Transition Manager" in detail, its functions, and how it works. Belatedly, yet importantly, we showcased our latest work on Cultural Heritage, an application for the Industrial Museum and Cultural Heritage of Thessaloniki [6], and viewed the results of our manager tested on this application.

## 4.2 Evaluation

To examine our system's overall user experience, we conducted a preliminary user-based evaluation with ten users (eight of them were proficient Unity3D developers, whereas the other two were relatively new to Unity3D). The main research questions were the following:

- How time-consuming is the manual SDK installation/reality switch/SDK uninstallation for a project in Unity3D?
- Is the SDK download function of our system preferable in comparison to the manual downloading procedure?
- Is the "Switch Reality" feature of our system preferable in comparison to the manual platform switch and settings application procedure?
- Is the SDK Uninstallation feature of our system preferable in comparison to the manual deletion of each SDK?
- Is the "Spawn Current Reality Camera" function of our manager useful?
- Would a developer prefer to use our manager for their Unity3D projects?

### 4.2.1 Methodology and participants

We divided the experiment into four different parts, each for one of the questions previously described. It is worth noting that all the participants were software developers, and they were familiar with Unity3D.

#### 4.2.1.1 Part 1: Manual Installation of ARCore SDK

In this part, we asked the participants to manually download and install an SDK (specifically ARCore SDK). They had to do it manually at first. They had to search for tutorials (in case they did not know how to install it). To succeed in this part, they had to include all the necessary files and directories of the SDK needed to compile successfully (thus, no compile errors in the Unity3D editor).

#### 4.2.1.2 Part 2: XR Transition Testing

In this part, we asked the participants to switch to the reality supported by the SDK they downloaded in the previous step. Again, they had to do it manually. To succeed in this step, they had to press the "Play" button of the Unity3D editor to run without any errors. For this part, we measured both the time they needed to initiate a platform switch manually and apply all the necessary settings. Finally, we received their feedback on the intricacy of the procedure.

#### 4.2.1.3 Part 3: SDK Uninstallation

In this part, we asked the participants to uninstall the SDK they installed in the previous steps. To succeed in this step, they had to reverse the Unity3D editor to its initial state (when they started the evaluation). They had to do it manually. In the end, we measured the time they needed and received their feedback regarding the complexity of the procedure.

#### 4.2.1.4 Part 4: Performing the tasks with "XR Transition Manager"

As this evaluation's final step, we asked the participants to perform the previous actions again, but this time, using our manager. This procedure was speedy (ten to twenty seconds) for users to find and click the appropriate buttons on the "Realities" menu. We should note that the SDK download time and system execution time for platform switch were not calculated, since that depends on the system and internet connection speed. In the end, we asked them to provide us with their feedback about the current procedure, and finally, we asked them if they would prefer to use our "XR Transition Manager" in their future Unity3D projects.

### 4.2.2 Results

#### 4.2.2.1 Part 1: Manual Installation of ARCore SDK

For this part, we measured the average time they needed to install ARCore SDK to their Unity3D project. We told them to search the internet freely for tutorials and steps on how to do this. For this part, time was our variable. It is important to note that we did not measure the time needed for downloading the SDK for this step, or the SDK system installation procedure. We present the results for this part in the table below.

| Participant | Hours | Minutes | Seconds | Time Decrease Percentage (for average time of 15 s) |
|---|---|---|---|---|
| #1 | 0 | 20 | 0 | 98.75% |
| #2 | 0 | 8 | 10 | 96.94% |
| #3 | 0 | 9 | 55 | 97.48% |
| #4 | 0 | 5 | 8 | 95.13% |
| #5 | 0 | 9 | 40 | 97.41% |
| #6 | 0 | 8 | 33 | 97.08% |
| #7 | 0 | 7 | 22 | 96.61% |
| #8 | 0 | 5 | 47 | 95.68% |
| #9 | 0 | 17 | 9 | 98.54% |
| #10 | 0 | 7 | 36 | 96.71% |

**Table 2.** Participants' results for the first part of our evaluation.

#### 4.2.2.2 Part 2: XR Transition Testing

For this part, we also measured the average time the participants needed to manually perform a transition across XR and prepare a sample scene of ARCore to run in Unity3D editor. Again, the participants were free to search for tutorials online on how to do this, in case they did not know. We did not measure the time needed for platform switch, since it depends on the system hardware. We present the results in the table below.

| Participant | Hours | Minutes | Seconds | Time Decrease Percentage (for average time of 15 s) |
|---|---|---|---|---|
| #1 | 0 | 10 | 0 | 97.50% |
| #2 | 0 | 5 | 0 | 95.00% |
| #3 | 0 | 1 | 10 | 78.57% |
| #4 | 0 | 2 | 16 | 88.97% |
| #5 | 0 | 4 | 43 | 94.70% |
| #6 | 0 | 12 | 31 | 98.00% |
| #7 | 0 | 18 | 27 | 98.64% |
| #8 | 0 | 8 | 23 | 97.02% |
| #9 | 0 | 22 | 34 | 98.89% |
| #10 | 0 | 14 | 32 | 98.28% |

**Table 3.** Participants' results for the second part of our evaluation.

#### 4.2.2.3 Part 3: SDK Uninstallation

It is the final part where we measured time. In this part, we wrote down the time the participants needed to remove ARCore SDK from their Unity3D project and return the project to its initial state. Again, system-dependent time was not measured (platform switch). We present the results in the following table.

| Participant | Hours | Minutes | Seconds | Time Decrease Percentage (for average time of 15 s) |
|---|---|---|---|---|
| #1 | 0 | 5 | 21 | 95.33% |
| #2 | 0 | 2 | 7 | 88.19% |
| #3 | 0 | 0 | 40 | 62.50% |
| #4 | 0 | 4 | 6 | 93.90% |
| #5 | 0 | 7 | 11 | 96.52% |
| #6 | 0 | 1 | 58 | 87.29% |
| #7 | 0 | 2 | 38 | 90.51% |
| #8 | 0 | 1 | 35 | 84.21% |
| #9 | 0 | 4 | 31 | 94.46% |
| #10 | 0 | 2 | 14 | 88.81% |

**Table 4.** Participants' results for the third part of our evaluation.

#### 4.2.2.4 Part 4: Performing the tasks with "XR Transition Manager".

After participants finished doing the tasks manually, we requested to perform them again using the "XR Transition Manager". It was a relatively quick procedure because they execute these actions with a button press. So that would be a few seconds (ten to twenty) for each participant. Finally, we asked the participants to give us their feedback by answering some questions.

### 4.2.3 Discussion

The current evaluation brought us fascinating results. Some participants were quick enough for both installation and transitioning across XR as well, whereas some participants took their time (about twenty minutes). For a programmer's development, twenty minutes is quite a long time for such a procedure, which our manager seeks to avoid. The same applies to the XR transition procedure as well (twenty-two minutes was the longest). The uninstallation procedure was quicker, although some cases needed some time (five or seven minutes, for instance, is considered quite long for such a case).

Then, participants tried our manager. The amount of time needed for all these tasks was between ten to twenty seconds per task for everyone. By taking the average of this time (15 seconds), we are able to calculate a percentage of saved time for each participant.

We calculated the time decrease percentage (in seconds) for each participant and for each evaluation step. We calculated each percentage using the following formula:

$$\% = \frac{ManualTime - AverageTransitionFrameworkTime}{ManualTime * 100}$$

"Manual Time" stands for the time (in seconds) that participants needed to do a requested action without using our framework. "Average Transition Framework Time" stands for the average time needed for participants to do an action by using our framework. This "time" is equal to 15 seconds, the average of a ten to twenty seconds window that each participant needed. These results are vital because our framework managed to save between 62.50% (worst case) and 98.89% of the participants' time (best case). Conclusively, our framework saved more than 62.50% of the participants' time (above average), which is clearly a win for us. Our current aim is to raise the worst-case percentage even more, in order to benefit even more developers.

Apart from the time-consuming scenario, we also wished to know our participants' thoughts about our manager and, more importantly, its ease of utilization. By observing Fig. 12, we understand that all participants gave a score of seven-plus out of ten for the SDK installing and XR transition tasks. For the uninstallation procedure, participants graded our manager with eight and nine out of ten. Two participants graded it with a six. We believe that is because our manager's SDK uninstallation procedure does not handle the cases where a plugin (.dll) might be used from Unity3D when users try to uninstall the SDK. This prevents our manager from entirely removing the SDK. It is a challenge for us to solve in the future.

The majority of participants delivered a ten to our "Spawn Current Reality Camera function" . Two participants gave it a seven out of ten, and one participant gave it a six. Belatedly, the participants seemed to approve our manager in general, by stating that they would use it in their Unity3D projects. They all gave a score of eight-plus out of ten when they were asked this question.

### 4.2.4 Evaluation Conclusion

To conclude the evaluation, the participants appeared to approve our framework, which made us more eager to upgrade it in the future continually. Our manager received more than a 60% score for each participant's feedback, which is quite a positive result. Besides, our manager saved more than 62.50% of their spent time. It is why we intend to work eagerly for the next months to stabilize it further and add more abilities that will be beneficial for developers. We present these additions in the following section.

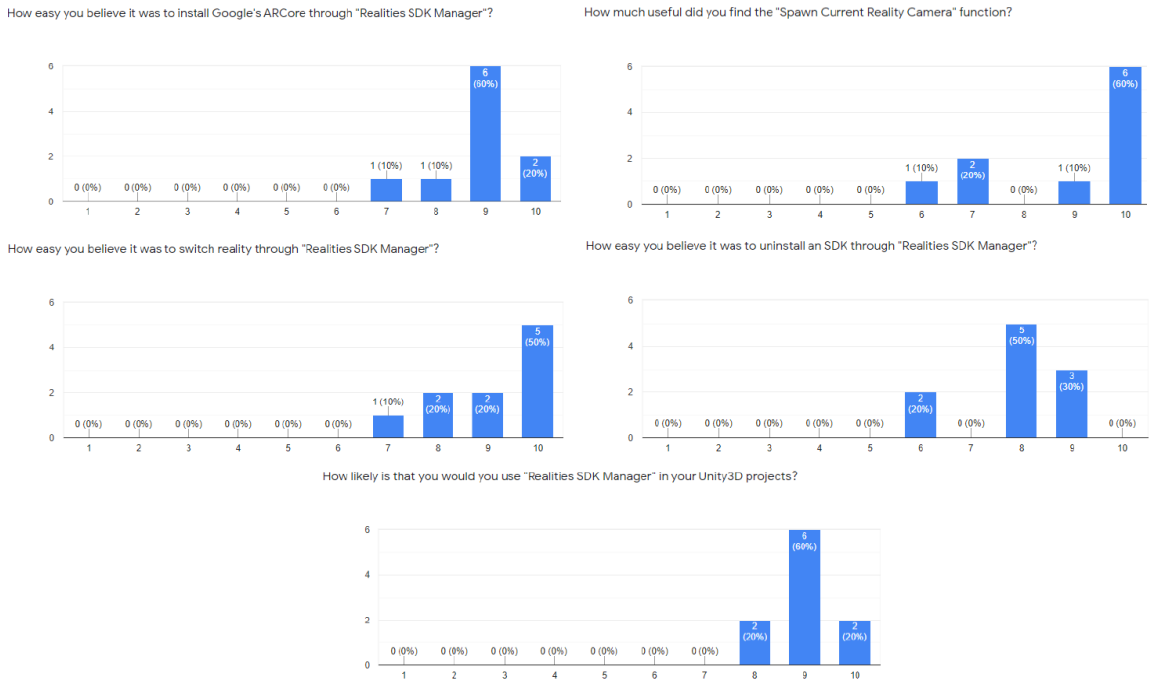

**Figure 12.** The results of the evaluation, depicted using graphs.

## 5 Future Work

We wanted to add some additions/features to our framework but could not due to the lack of time. We believe that these features would make the "XR Transition Manager" even more helpful to developers. We divided these features into the following sections.

### 5.1.1 Saving Time

Saving time is one of the most critical elements of our manager, if not the most prominent. It already saves a reasonable amount of time by managing four different SDKs and transcending across XR based on them.

Until now, our manager successfully substitutes the project camera (which is selected by the developer). It adds all the necessary objects in the scene required by the current SDK to work correctly. It also applies all the necessary settings and installs the packages needed. However, we do nothing regarding the controls of the application. For instance, suppose we would like to transform an application operating on Oculus Go (mobile virtual reality) to run on a mobile device with Augmented Reality. Oculus Go has a controller, with which users can perform various interactions in their applications. Suppose an application moves to Augmented Reality (from Virtual Reality) by utilizing our manager. In that case, users would not be able to perform any interactions since such mobile devices use completely different ways of interaction. For instance, most mobile devices use the touch screen to receive input from users. We want to provide our manager with a way to map all the supported functions from the input module of the originating reality to the target reality. For example, the pressing of a button in the Oculus Go controller could be mapped to a simple touch on a mobile device's touchscreen or a pinch gesture for HoloLens.

### 5.1.2 Expanding availability

We want to benefit as more developers as possible with this framework. At its current state, the number of developers who can profit is somehow limited. For developers to use this framework, they must be familiar with Unity3D, and their target device for their application must be a mobile device, Oculus Go, or Microsoft HoloLens. We want to put an end to this boundary by expanding the number of game engines on which our manager will be installed and the number of supported devices and SDKs. As a result, our foremost future objective is to make our manager available for the Unreal game engine since many developers use it. Moreover, we would like to provide support to all desktop Virtual Reality devices, SDKs, and mobile headsets/devices. That would significantly increase the number of developers who can be benefited by our manager. Last but not least, we would like to adjust the manager properly, like converting it to a single package or DLL file, to distribute way more efficiently.

### 5.1.3 Complete reality transformation

Our framework may currently transit a 3D application across XR, but it does not consider any theoretical elements that the target reality may have. We want to add this feature in the future. This feature will scan the application scene's contents and perform some necessary changes regarding the target reality. For instance, if we move an Augmented Reality application containing a portal to Virtual Reality, that portal would be of no use. That is because Virtual Reality provides users with full immersion, and thus they do not have access to the real world. In that case, our framework should be able to detect such portals and disable them.

### 5.1.4 Smart Performance Adaptation

Our manager's main objective is to make the application porting procedure easier for developers, so that eventually, in the future, more and more 3D applications will be available for many devices. All those devices are different, though, in terms of hardware and overall performance. It is something that we would also like our manager to consider. We describe this in detail below.

#### 5.1.4.1 Complexity of the models

The 3D scenes of such projects contain different kinds and sizes of 3D models. Sometimes, 3D models tend to be very complicated because they might contain many vertices or complicated geometry. The more complex a model is, the less likely the application will operate smoothly on devices with lower specifications. That is a challenge that we would like to relieve the developers from. We plan to upgrade our manager to consider the device that the application will operate on and apply the necessary quality settings for the application to run smoothly on that device. We will also examine the case of applying reduction algorithms to those meshes, to diminish the geometry complexity and make them lighter for low specs devices.

#### 5.1.4.2 Lighting

One common solution to overcome low performance, concerning lighting, especially in low specs devices, is static lighting. It makes the scene less realistic but increases performance. We want to add some global illumination calculation algorithms to our manager. Again, depending on the target device, if the device is a low specs one, our manager would provide light and realistic real-time illumination algorithms. Even in those devices, real-time lighting will be available. We believe that such a feature will significantly improve the application quality while operating on low specs devices and render them highly realistic.